\documentclass[sigconf, nonacm]{acmart}

\usepackage[utf8]{inputenc} 
\usepackage[T1]{fontenc}    
\usepackage{hyperref}       
\usepackage{url}            
\usepackage{booktabs}       
\usepackage{amsfonts}       
\usepackage{nicefrac}       
\usepackage{microtype}      
\usepackage{amsmath}
\usepackage{amsthm}
\usepackage{enumitem}
\usepackage{mdframed}
\usepackage{graphicx}
\usepackage{adjustbox}
\usepackage{cleveref}
\usepackage{setspace}
\usepackage{subcaption} 
\usepackage{fancyvrb}

\usepackage{booktabs}
\usepackage{caption}
\usepackage{siunitx}

\theoremstyle{definition}
\newtheorem{example}{Example}
\newcommand{\stitle}[1]{\vspace{2pt}\noindent\textbf{#1}}

\usepackage{color}

\title{Data Cleaning Using Large Language Models}

\author{Shuo Zhang}
\email{sz3177@columbia.edu}
\affiliation{
  \institution{Columbia University}
}

\author{Zezhou Huang}
\email{zh2408@columbia.edu}
\affiliation{
  \institution{Columbia University}
}

\author{Eugene Wu}
\email{ewu@cs.columbia.edu}
\affiliation{
  \institution{DSI, Columbia University}
}

\begin{document}

\begin{abstract}

Data cleaning is a crucial yet challenging task in data analysis, often requiring significant manual effort. To automate data cleaning, previous systems have relied on statistical rules derived from erroneous data, resulting in low accuracy and recall. This work introduces Cocoon, a novel data cleaning system that leverages large language models for rules based on semantic understanding and combines them with statistical error detection.
However, data cleaning is still too complex a task for current LLMs to handle in one shot. To address this, we introduce Cocoon, which decomposes complex cleaning tasks into manageable components in a workflow that mimics human cleaning processes.
Our experiments show that Cocoon outperforms state-of-the-art data cleaning systems on standard benchmarks.

\end{abstract}

\maketitle

\section{Introduction}

Data cleaning is a well-known, challenging yet crucial task. Datasets frequently contain extreme or erroneous values, which can greatly affect outcomes of downstream analytics~\cite{sambasivan2021everyone, huang2023data}. Consequently, it's well known that analysts spend over 80\% of their time manually reviewing and cleaning data~\cite{eckerson2002data}. To tackle these challenges, previous systems aim at automating the data cleaning process ~\cite{chu2013holistic,rekatsinas2017holoclean,mahdavi2020baran}.

However, traditional data cleaning methods struggle with low accuracy and low recall because they rely on statistical rules like thresholds, distributions, dependencies, denial constraints, etc., to detect anomalies ~\cite{mayfield2010statistical,beskales2010fd,rekatsinas2017holoclean}. The problem is that \textbf{these detection and cleaning rules are derived statistically} from unreliable, erroneous data, and therefore have low quality. On the other hand, data reflects real-world entities. When humans manually clean data, they can \textbf{semantically detect these rules and propose cleaning strategies} based on external real-world knowledge, resulting in much better performance. To demonstrate the limits of statistical rules and the importance of semantic knowledge, consider the cleaning in the Rayyan table~\cite{ouzzani2016rayyan} as an example:

\begin{example} Consider the data cleaning process for the 'article\_language' column in Rayyan. The \textbf{statistical detection} analyzes the distribution of each unique value. It reveals that 46.4\% of entries are "eng" and 9.5\% are "English". As these strings don't exhibit strong distribution or pattern outliers, no \textbf{statistical error} is detected by past detection systems including Holoclean~\cite{rekatsinas2017holoclean} and Baran~\cite{mahdavi2020baran}. However, when analysts manually clean the data, they \textbf{detect semantically} that "eng" and "English" are redundant representations of the same concept of english language.  The \textbf{semantic cleaning} process cleans "English" $\to$ "eng" because "eng" is the most common representation, and similarly cleans other language values like "French" $\to$ "fre", "German" $\to$ "ger", and "Chinese" $\to$ "chi" in the 'article\_language' column.
\end{example}

Such a process of applying semantic understanding of the tables and values for data detection and cleaning extends beyond just the data quality issue of inconsistent value representations. \Cref{sec:taskdecom} catalogs the various data quality issues studied, and the detailed detection and cleaning steps.
 
To provide such semantic understanding without much human effort, recent advancements in large language models, such as GPT-4 and Claude 3.5, have demonstrated near-human-level general capabilities across various tasks~\cite{bubeck2023sparks}. However, effectively prompting these models for data cleaning purposes remains a challenge due to the complex and ambiguous nature of data cleaning tasks. Our experimental results indicate that existing data cleaning tools~\cite{ahmad2023retclean,qi2024cleanagent} utilizing LLMs achieve close to zero accuracy and recall for the majority of standard data cleaning benchmarks.

\begin{figure}
    \centering

        \centering
        \begin{subfigure}[b]{0.8\linewidth}
            \centering
            \includegraphics[width=0.5\linewidth]{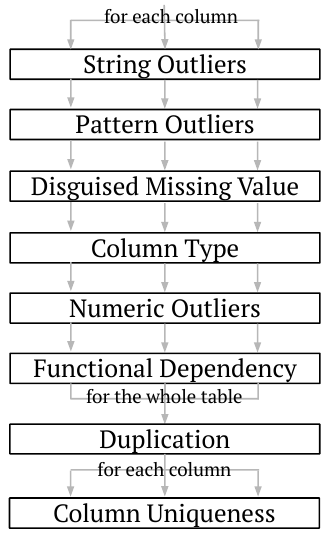}
            \caption{}
            \label{fig:issue_decom}
        \end{subfigure}
        \hfill
        \begin{subfigure}[b]{0.8\linewidth}
            \centering
            \includegraphics[width=\linewidth]{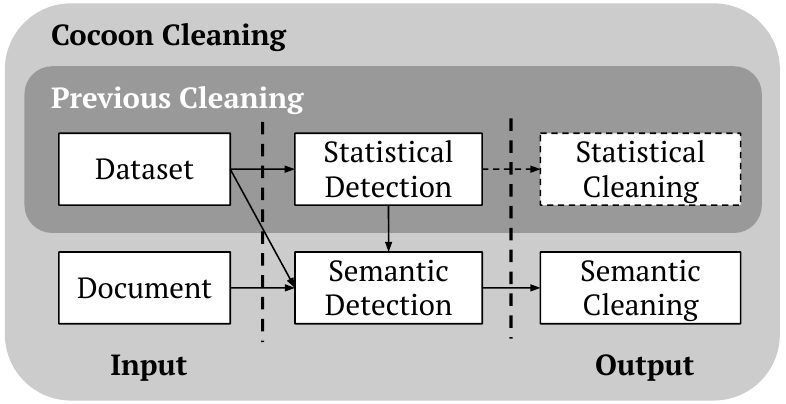}
            \caption{}
            \label{fig:task_decom}
        \end{subfigure}

    \caption{Cocoon decomposes data cleaning in two dimensions: 
             (a) decompose it for different types of errors, for each column; 
             (b) For each type of error, Cocoon decomposes the cleaning steps 
             with traditional statistical detection, combined with semantic 
             error detection and cleaning.}
    \label{fig:process}
\end{figure}

To this end, we introduce Cocoon\footnote{Open sourced at \url{https://cocoon-data-transformation.github.io/page/clean}}, a data cleaning system based on our previous data profiling system~\cite{huang2024cocoon} that leverages LLMs for semantic understanding and combines statistical error detection for better context. 
To tackle the complexities of data cleaning, our core approach is to decompose tasks into manageable components; such an approach is crucial for the accuracy and robustness of data tasks like visualization and transformation~\cite{khot2022decomposed, dibia2023lida, huang2024relation, huang2150transform}. To design the decomposition for data cleaning (\Cref{fig:process}), Cocoon mimics the cleaning steps a human would take: (1) we first breaks down the data cleaning into small common issues including duplication, missing values, and outliers~\cite{rahm2000data,chu2016data}, (2) then for each data cleaning issue, we breaks down the task into statistical detection, semantic detection, and semantic cleaning. 
Cocoon outperforms all other state of the art data cleaning systems on 4 out of 5 standard benchmarks, achieving higher F1 scores. For the one exception, we demonstrate that it is due to the ambiguity of the benchmark.

\section{System Design}

The challenge of applying LLMs to data cleaning is twofold: (1)  data is too large to fit into the prompt of current LLMs, and (2) the whole data cleaning task is too complex for LLMs to perform in a single pass. To address (1), Cocoon leverages traditional statistical methods to profile the tables~\cite{kandel2012profiler,epperson2023dead} (e.g., value distribution, missing percentages) and includes these in the prompt to help LLMs better understand the data. To tackle (2), Cocoon decomposes data cleaning into separate processes for different types of errors (illustrated in \Cref{fig:process}), motivated by how human decomposed data cleaning errors~\cite{rahm2000data,chu2016data}. We begin by discussing the decomposition, followed by the implementation details.

\subsection{Task Decomposition}
\label{sec:taskdecom}
Cocoon detects and cleans the following types of data quality issues:

\subsubsection{String Outliers}
\begin{figure}[ht]
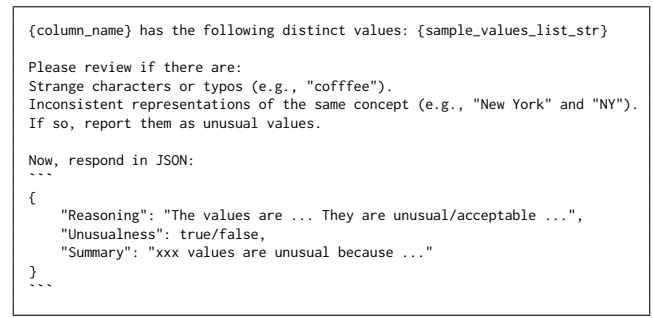

\begin{Verbatim}[fontsize=\scriptsize, frame=single, framesep=2mm]
{column_name} has the following distinct values: {sample_values_list_str}

Please review if there are:
Strange characters or typos (e.g., "cofffee"). 
Inconsistent representations of the same concept (e.g., "New York" and "NY").
If so, report them as unusual values.

Now, respond in JSON:
```
{
    "Reasoning": "The values are ... They are unusual/acceptable ...",
    "Unusualness": true/false,
    "Summary": "xxx values are unusual because ..."
}
```
\end{Verbatim}
    \caption{Prompt for Semantic Detection of string outliers for one column through samples.}
    \label{our_prompt}
\end{figure}

We sample frequent values (by default 1000) and let LLMs review whether these values semantically contain typos or inconsistent representations (prompt shown in \Cref{our_prompt}). If errors are found, we ask LLMs to build a mapping from erroneous to correct values, and execute the cleaning through \verb|CASE WHEN| clauses. To avoid run out of context for large datasets, we set the value batch size (by default 1000) and let LLMs evaluate one batch at a time (prompt shown in \Cref{our_prompt2}).

\subsubsection{Pattern Outliers}
We recursively ask LLMs to write a list of semantically meaningful regular expression patterns that cover all column values (e.g., \verb|\d{2}/\d{2}/\d{4}| for dates is meaningful based on the day/month/year, but \verb|.*| is not), and verify them with SQL. We then ask LLMs to assess these if there are inconsistent representations. Cleaning is via regex transformation.

\subsubsection{Disguised Missing Value (DMV)}
We show the column values and ask LLMs to identify values that are currently not \verb|NULL|, but semantically means that the value are missing (e.g., string values like "N/A", "null"). Cleaning is performed using a \verb|CASE WHEN THEN NULL| clause.

\subsubsection{Column Type}
We identify the current column type from the database catalog and ask LLMs to suggest the most suitable data type semantically. For cleaning, we use \verb|CAST| clauses.

\subsubsection{Numeric Outliers}
We capture the minimum and maximum values statistically and review the acceptable range semantically. We address outliers using a \verb|CASE WHEN| clause for thresholding.

\subsubsection{Functional Dependency}
Following Baran, we only consider FDs where both left and right-hand sides have a single attribute. We compute the entropy measurement of each FD pair~\cite{beskales2010fd}, and let LLMs review if these statistically strong functional dependencies are meaningful semantically. If meaningful, we identify all groups of values violating the functional dependency, ask LLMs to provide the correct mapping, and clean the data using a \verb|CASE WHEN| clause.

\subsubsection{Duplication}
The statistical error detection selects duplicated rows. If there are duplicates, we use LLMs to determine if these duplications are semantically acceptable (e.g., duplication in logging with coarse time granularity). If it's erroneous, cleaning is performed by \verb|SELECT DISTINCT|.

\subsubsection{Column Uniqueness}
Some columns, e.g., primary key, should be unique. We compute the unique ratio of each column statistically and ask LLM to decide if the column should be unique semantically. For cleaning, we ask LLM to build a window function keyed on the relevant column, assuming some column contains information to prioritize records (e.g., the latest time).

Note that the order in which data quality issues are addressed is important. Consider a column of dates that are human-entered with data quality issues. We need to start with string outlier detection to fix typo issues (e.g., "1/1/2000x" to "1/1/2000"). Only after typos are fixed can we detect the patterns (e.g., \verb|\d{2}/\d{2}/\d{4}|). And only after the patterns are standardized can we perform the CAST operation. Only when the column is cast to DATE can we show the distribution for numeric outliers.

\begin{figure}[ht]
\begin{Verbatim}[fontsize=\scriptsize, frame=single, framesep=2mm]
{column_name} is unusual: {summary}
It has the following values: {batch_values_list_str}

Maps those unusual values to the correct ones to address the problems.
If old values are meaningless, map to empty string.

Return in the following format:
```yml
explanation: >
    The problem is ... The correct values are ...
mapping:
    old_value: new_value 
```
\end{Verbatim}
    \caption{Prompt for Semantic Cleaning of string outliers for one column.}
    \label{our_prompt2}
\end{figure}

\subsection{Implementation}
Cocoon is implemented as a Python Library that connects to databases, and LLM APIs.
Cocoon supports common databases including Snowflake, DuckDB, BigQuery, and SQL Server\footnote{Pattern outliers are not supported in SQL Server due to its limited pattern matching capabilities.}. We support LLM APIs from Anthropic, Azure, Bedrock, VertexAI, and OpenAI.

To ensure that the error detection and cleaning processes are scalable, interpretable, and reusable, we perform them using SQL queries. The final output is a set of well-commented SQL queries.

Cocoon is designed to be a human-in-the-loop process for user feedback. For each error detection and data cleaning step, we present the LLM reasoning and ask humans to verify and adjust (details in \Cref{sec:ui}). In future work, such cleaning processes can take domain-specific documents as context.

\section{Experiment}
\label{sec:exp}

\begin{table*}[t]
\centering
\caption{Data cleaning performance (\textbf{P}recision, \textbf{R}ecall, and \textbf{F}-1) across different benchmarks. * Movies (4.6MB) is the largest dataset, and Holoclean runs out of memory, while CleanAgent doesn't accept files $>$2MB. Therefore, we benchmark them over the sample of the first 1000 rows.}
\label{tab:comparison}
\resizebox{\textwidth}{!}{
\begin{tabular}{l*{15}{c}}
\toprule
\textbf{System} & \multicolumn{3}{c}{\textbf{Hospital}} & \multicolumn{3}{c}{\textbf{Flights}} & \multicolumn{3}{c}{\textbf{Beers}} & \multicolumn{3}{c}{\textbf{Rayyan}} & \multicolumn{3}{c}{\textbf{Movies}} \\
\cmidrule(lr){2-4} \cmidrule(lr){5-7} \cmidrule(lr){8-10} \cmidrule(lr){11-13} \cmidrule(lr){14-16}
& P & R & F & P & R & F & P & R & F & P & R & F & P & R & F \\
\midrule
HoloClean & \textbf{1.00} & 0.46 & 0.63 & 0.73 & 0.34 & 0.47 & 0.05 & 0.04 & 0.04 & 0.53 & 0.67 & 0.59 & 0.00* & 0.00* & 0.00* \\
Raha+Baran & 0.91 & 0.60 & 0.72 & 0.84 & \textbf{0.61} & \textbf{0.70} & 0.97 & \textbf{0.96} & 0.96 & 0.83 & 0.35 & 0.50 & 0.85 & 0.75 & 0.80 \\
CleanAgent & 0.00 & 0.00 & 0.00 & 0.00 & 0.00 & 0.00 & 0.00 & 0.00 & 0.00 & 0.00 & 0.00 & 0.00 & 0.00* & 0.00* & 0.00* \\
RetClean & 0.00 & 0.00 & 0.00 & 0.00 & 0.00 & 0.00 & 0.00 & 0.00 & 0.00 & 0.52 & 0.48 & 0.50 & 0.00 & 0.00 & 0.00 \\
Cocoon & 0.87 & \textbf{0.93} & \textbf{0.90} & \textbf{0.91} & 0.42 & 0.57 & \textbf{0.99} & \textbf{0.96} & \textbf{0.97} & \textbf{0.88} & \textbf{0.84} & \textbf{0.86} & \textbf{0.91} & \textbf{0.83} & \textbf{0.87} \\
\bottomrule
\end{tabular}
}
\end{table*}

To evaluate the performance of Cocoon, we conducted experiments on 5 standard benchmarks.

\subsection{Experiment Setup}

All our experiments run on a Ubuntu 20.04 LTS machine with 16 virtual CPUs and 104 GB RAM. For database, We run DuckDB. For LLMs, we use Claude 3.5.

\stitle{Datasets.} We use 5 standard benchmarks representing a variety of domains and data quality issues:
\begin{itemize}[leftmargin=0.5cm]
    \item \textbf{Hospital} and \textbf{Flights} \cite{rekatsinas2017holoclean}: These datasets contain a variety of errors including typos, functional dependency violations, wrong column types and DMV.
    \item \textbf{Beers} \cite{mahdavi2019raha}: This dataset includes functional dependency errors and column type errors.
    \item \textbf{Rayyan} \cite{ouzzani2016rayyan} and \textbf{Movies} \cite{magellandata}: Besides column types and DMV, these real-world datasets contain many value misplacement errors like the county was incorrectly entered in the city column.
\end{itemize}

\stitle{Baselines.} We compare Cocoon against 4 baselines.
\begin{itemize}[leftmargin=0.5cm]
    \item \textbf{Cocoon.} Cocoon takes the database as input and outputs SQL. While the detection and cleaning process is intended to be HIL, we skip these and use the LLM provided ground truth.

    \item \textbf{Holoclean \cite{rekatsinas2017holoclean}.} Holoclean additionally takes denial constraints as input, for which we provide the ground truth. However, it runs out of memory on large datasets (Movies), so we use samples of the first 1000 rows.
    
    \item \textbf{Raha~\cite{mahdavi2019raha} and Baran~\cite{mahdavi2020baran}.} Raha first detects errors, and Baran cleans them. \textbf{Note that Baran additionally requires feedback on 20 clean cells. We provide the ground truth.}
    
    \item \textbf{CleanAgent \cite{qi2024cleanagent}.} CleanAgent inputs and outputs CSV files, primarily for standardization.
    
    \item \textbf{RetClean \cite{ahmad2023retclean}.} RetClean can accept additional tables as inputs, but we do not have any to provide.
    
\end{itemize}

\textbf{Evaluation.} Current benchmarks (1) are ambiguous and (2) don't handle column types and DMV:
\begin{itemize}[leftmargin=0.5cm]
\item \textbf{Case Sensitivity}: Different cases are acceptable as long as the case is consistent across values.

\item \textbf{Column Type}: Previous data cleaning systems, like our baselines, are limited because they operate on CSV files without rich column types. However, in databases, certain values, such as "yes"/"no," are better represented as bool. Cocoon casts them to "True"/"False", but for other data cleaning systems, we consider them correct even if they do not perform these casts.

\item \textbf{DMV}: No baseline system casts DMV (e.g., "N/A") to NULL, but we still consider them correct.
\end{itemize}

We don't consider these, but Cocoon shows better results when accounting for them (\Cref{sec:erroranalysis}).

\subsection{Results}

\stitle{Performance.}
Table \ref{tab:comparison} reports the precision, recall and F1. Cocoon outperforms all four baseline systems in terms of F1-scores for all but the Flights dataset. 

\begin{itemize}[leftmargin=0.5cm]
\item \textbf{Flights Benchmark Ambiguity}: 
Cocoon achieves high precision but low recall for the Flights dataset due to benchmark ambiguity. This is because of the ambiguous FD:  Flight Number $\to$ Actual Departure/Arrival Time. This FD is ambiguous because the original dataset frequently contains inconsistent departure and arrival times. For example, the actual arrival time for flight "AA-1733-ORD-PHX" is recorded as "10:30 p.m." in 5 rows, "10:31 p.m." in 4 rows, "10:28 p.m." in 3 rows, "10:39 p.m." in 1 row. It's nearly impossible to determine the true arrival time and clean the data accurately.
Furthermore, such inconsistencies appear to be application issues rather than data cleaning issues, making it preferable to preserve these to represent the uncertainty.

\item \textbf{HoloClean} performs poorly because its error detection relies heavily on integrity constraints provided by the user. Despite the provided ground truth constraints, most inconsistency issues (e.g., "oz" vs. "ounce" in Beers, and "100 min" vs. "1 hour 40 min" in Movies) cannot be adequately captured by these constraints.
We notice that the recall for the Hospital dataset is lower than the results reported in the original HoloClean paper. Despite experimenting with various threshold values ($\tau$), we were unable to replicate their results. Nevertheless, Cocoon outperforms HoloClean in terms of recall and F1 score provided in the original Holoclean paper (0.713 and 0.832, respectively).

\item \textbf{CleanAgent \& RetClean} utilize LLMs to clean data. However, CleanAgent achieves low results as it focuses on standardizing categories (e.g., email, phone, date). RetClean primarily cleans tables using external clean tables, which are not available. It only performs well on Rayyan because Rayyan contains a large number of typos obvious for LLMs to fix.

\item \textbf{Raha \& Baran.} These demonstrate reasonable performance across all benchmarks. However, they use traditional ML models (e.g., Gradient Boosting, Adaboost), which are much less capable and lack the semantic understanding ability, compared to LLMs with billions of parameters. They also additionally require ground truth values provided by users. 

\end{itemize}

\section{Conclusion}
In this work, we introduced Cocoon, a data cleaning system that leverages large language models to enhance semantic understanding in the data cleaning process. By decomposing complex data cleaning tasks into multi-step components that mimic human cleaning styles, Cocoon outperforms all other SOTA cleaning systems on 4 out of 5  benchmarks. Future work will address domain specific errors by exploring autonomous error classification and agent-based approaches.



\bibliographystyle{plain}
\bibliography{main}

\appendix

\section{User Interface}
\label{sec:ui}
Cocoon is designed to be interpretable using the HIL process with NL explanations from LLMs. Specifically, we ask users to provide feedback on (1) the semantic detection results of error types, and (2) the semantic cleaning that maps incorrect values to clean values (see \Cref{fig:process}).

\begin{figure*}[htbp]
    \centering
    \setlength{\fboxsep}{0pt} 
    \setlength{\fboxrule}{0.5pt} 
    \fbox{\includegraphics[width=\linewidth]{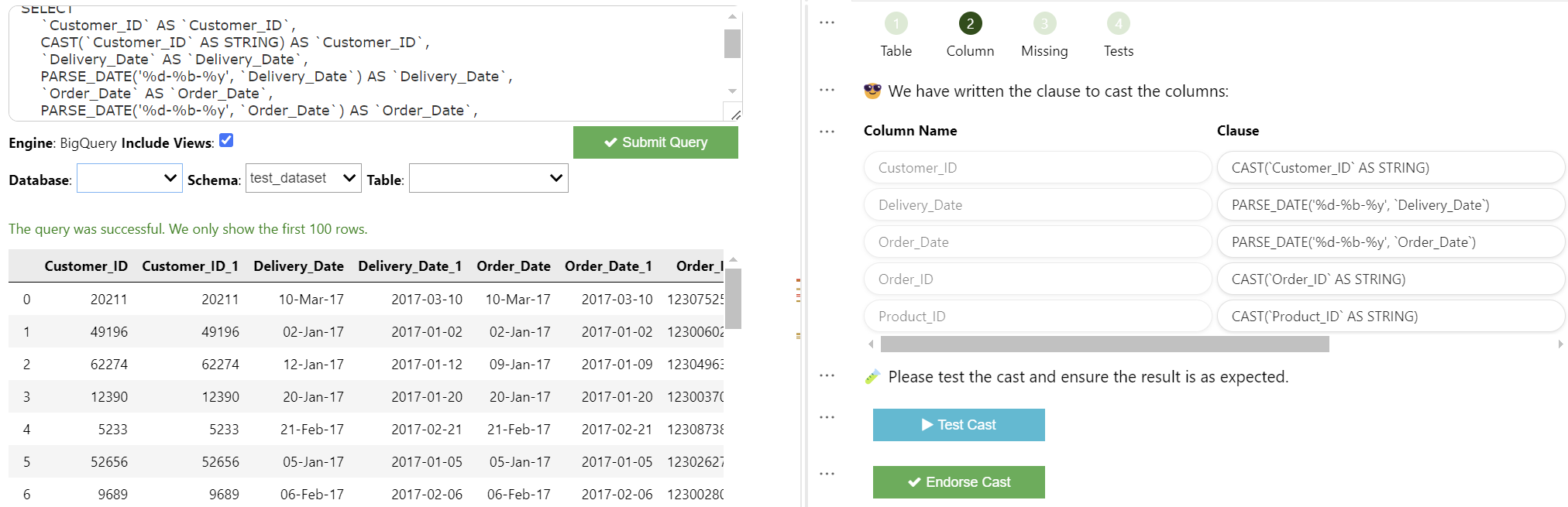}}
    \caption{The UI for each data cleaning step. The right side is the interface where users specify the SQL clauses for column cast. The left side is the query interface to preview the results.}
    \label{fig:ui-clean-step}
\end{figure*}

\begin{figure*}[htbp]
    \centering
    \setlength{\fboxsep}{0pt} 
    \setlength{\fboxrule}{0.5pt} 
    \fbox{\includegraphics[width=\linewidth]{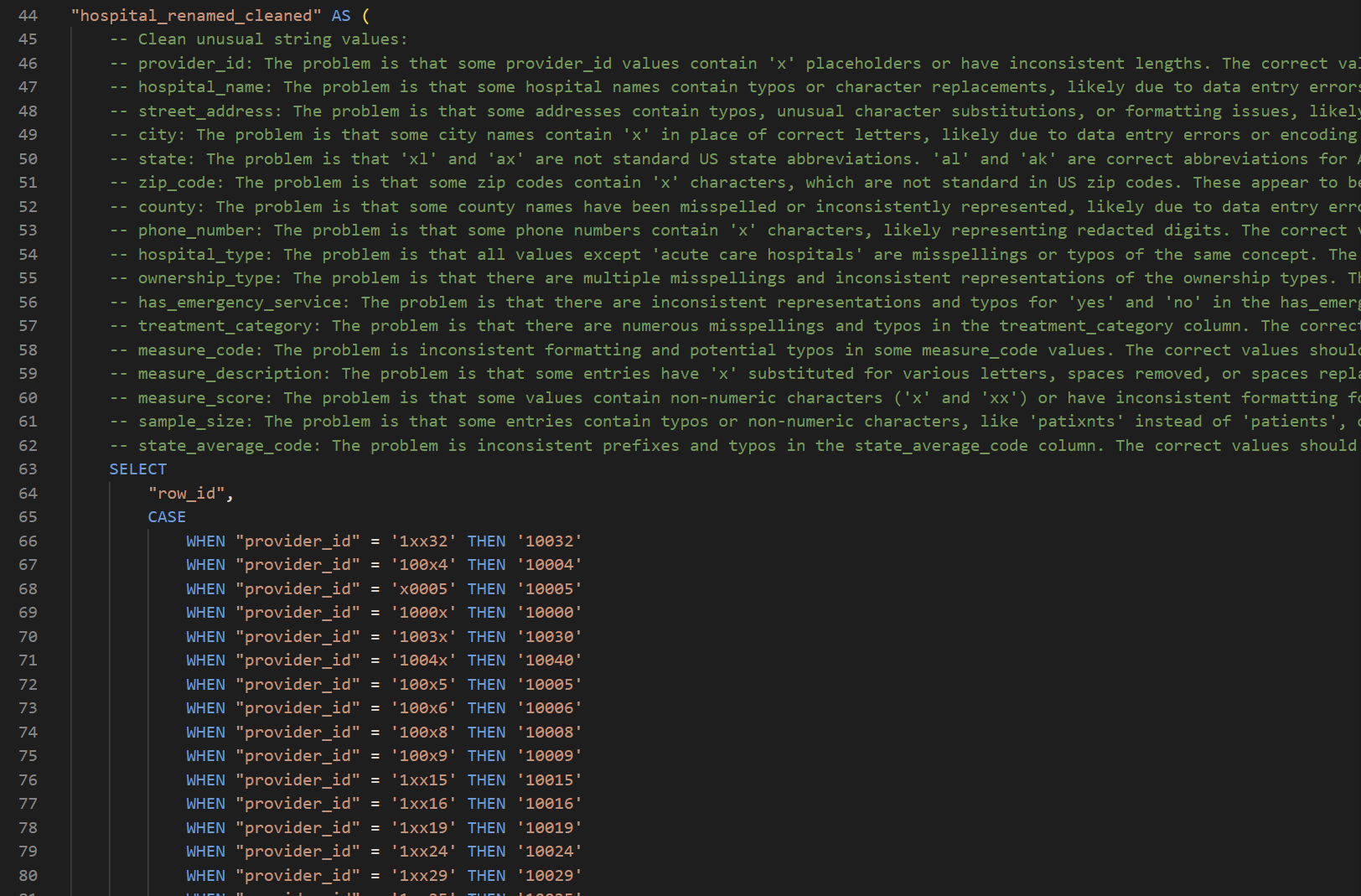}}
    \caption{Output SQL queries for results. We provide the cleaning reasoning as NL in the comments and use SQL for cleaning.}
    \label{fig:output-sql-queries}
\end{figure*}

Figure~\ref{fig:ui-clean-step} shows the interface where users can specify and view the SQL results for the semantic cleaning of column types.
The final output is aimed to be interpretable for long-term maintenance and scalable for existing data pipelines. We construct SQL queries documented with the reasoning behind the cleaning process. As shown in Figure~\ref{fig:output-sql-queries}, we provide examples of how to map values for string outliers, along with natural language descriptions.

Full illustrations of the cleaning results in HTML reports and commented SQL pipelines are available at \url{https://cocoon-data-transformation.github.io/page/clean}. In the future, we will conduct user studies to better understand how humans interact with Cocoon through natural language.

\begin{table*}[t]
\centering
\caption{Distribution of Various Types of Errors Across Benchmarks}
\label{table:error-summary}
\begin{tabular}{@{}lccccccl@{}}
\toprule
\textbf{Dataset} & \textbf{Size} & \textbf{Typo} & \textbf{FD} & \textbf{Column Type} & \textbf{Inconsistency} & \textbf{DMV} & \textbf{Misplacement} \\
\midrule
Hospital & $1000 \times 19$ & 213 & 331 & 3,000 & -- & 227 & -- \\
Movies & $7390 \times 17$ & 184 & -- & 14,433 & -- & 131 & 938 \\
\bottomrule
\end{tabular}
\end{table*}

\section{Error analysis}
\label{sec:erroranalysis}
In this section, we present the error type details in the benchmarks. Previous data cleaning work focused on errors such as typos, functional dependency violations, and misplacements. However, we find that there are other types of errors in the benchmarks that are disregarded, specifically forms of disguised missing values~\cite{qahtan2018fahes} and column type errors. \Cref{table:error-summary} shows the distribution of errors in the datasets of Hospital and Movies. Specifically, column type errors are very common. For example, for "EmergencyService" in the hospital dataset, the current values are "yes" and "no", which semantically means a boolean. Results in \Cref{sec:exp} don't consider these errors. Table \ref{tab:comparison_amb} presents the results of Cocoon and the baseline systems when these are considered as errors. Cocoon outperforms all 4 baselines with >0.9 F1 score. This outcome is anticipated, as the precision and recall increase when new errors are introduced, and Cocoon effectively corrects them. Among all 4 baselines, only Raha partially solves the column type casting, as it asks for the ground truth samples and fixes "yes/no" -> bool. However, Raha struggles with more complex transformations for fields with higher cardinality, and transformations that necessitate semantic understanding. For instance, it fails to consistently convert time expressions like "1 hr. 30 min." and "90 min" into float 90.

\begin{table}
\centering
\caption{Comparison with other data cleaning systems. }
\label{tab:comparison_amb}
\begin{tabular}{@{}lS[table-format=1.2]S[table-format=1.2]S[table-format=1.2]S[table-format=1.2]S[table-format=1.2]S[table-format=1.2]@{}}
\toprule
& \multicolumn{3}{c}{\textbf{Hospital}} & \multicolumn{3}{c}{\textbf{Movies}} \\
\cmidrule(lr){2-4} \cmidrule(l){5-7}
\textbf{Approach} & {P} & {R} & {F} & {P} & {R} & {F} \\
\midrule
HoloClean & 1.00 & 0.13 & 0.24 & 0.00 & 0.00 & 0.00 \\
Raha & 1.00 & 0.97 & 0.98 & 0.57 & 0.55 & 0.56 \\
CleanAgent & 0.00 & 0.00 & 0.00 & 0.00 & 0.00 & 0.00 \\
RetClean & 0.00 & 0.00 & 0.00 & 0.00 & 0.00 & 0.00 \\
Cocoon & \bfseries 0.99 & \bfseries 0.99 & \bfseries 0.99 & \bfseries 0.96 & \bfseries 0.91 & \bfseries 0.93 \\
\bottomrule
\end{tabular}
\end{table}






\end{document}